\documentclass[journal]{IEEEtran}
\usepackage{cite}
\ifCLASSINFOpdf
\else
\fi

\usepackage[cmex10]{amsmath}
\usepackage{array}
\usepackage{mdwmath}
\usepackage{mdwtab}
\usepackage{graphicx}
\usepackage{euscript}
\usepackage{amsfonts}
\usepackage{bigstrut}
\usepackage{tabularx}
\usepackage{subcaption}
\usepackage{cite}
\usepackage{amsmath}
\usepackage{algorithm}
\usepackage[noend]{algpseudocode}
\makeatletter
\def\BState{\State\hskip-\ALG@thistlm}
\makeatother
\usepackage{enumitem}
\usepackage{upgreek}
\usepackage{dsfont}
\usepackage{amsmath}
\usepackage{amsthm}
\usepackage{amssymb}
\usepackage{mathtools}

\usepackage{hyperref}
\usepackage{stfloats}
\usepackage{epstopdf} 
\usepackage{bbm}
\usepackage{xcolor}

\begin{document}

\title{Next Generation Mega Satellite Networks: \\Opportunities, Challenges, and Performance}

\author{Bassel Al Homssi, Akram~Al-Hourani, Ke~Wang, Phillip~Conder,\\ Sithamparanathan~Kandeepan, Jinho~Choi, Ben~Allen, and Ben~Moores.
}

\maketitle

\begin{abstract}

	Digital connectivity has become the foundation of prosperity and an essential need for functioning societies. Despite this dependence, limitation on Internet access remains a prevalent issue, largely hinged on socioeconomic and geographic factors. A promising solution to attain global access equality is based on integrated terrestrial-satellite networks that rely on low Earth orbit (LEO) mega constellations. While the benefits of LEO constellations complement the shortcomings of terrestrial networks, their incorporation impacts the network design, adding complexity and challenges. This article presents a systematic analysis of next generation LEO mega satellite constellations and outlines opportunities by virtue of the many benefits these constellations can provide and highlights the major challenges. Furthermore, it provides a synopsis of analytic models and underscores modern simulation approaches for next generation mega satellite constellations. This provides network designers with the necessary insights into satellite network performance.
\end{abstract}

\begin{IEEEkeywords}
	LEO constellations, Stochastic geometry, Beyond 5G, Access equality, Satellite communications.
\end{IEEEkeywords}

\IEEEpeerreviewmaketitle

\section{Introduction}
The contemporary version of modernity is founded on digitalization, computing, and ubiquitous Internet access. The far-reaching impact in which their elements touch our lives are lamented in our social interactions, modern economy, healthcare systems, and education. Started as a visionary approach to share information, Internet access is now a necessity to meeting basic human needs. However, Internet availability, reliability, and coverage, have not reached a truly global extent. The digital divide is a civil rights issue that more severely impacts underprivileged and disadvantaged communities. Therefore, there is an urging need to develop effective strategies and solutions that eliminate persisting digital divides, hence establishing fair and truly global coverage.

Current network infrastructure, wired or wireless, is dominated by terrestrial communication services, the deployment of which is largely driven by short-term profits, exacerbating the access inequality issue by under-connecting remote areas that lack sufficient business motives. Non-Terrestrial Networks (NTN) recently emerged as a promising solution to reinforce the connectivity of terrestrial networks by providing extended coverage. One of the main components in NTNs is satellite networks, including recent low Earth orbit (LEO) mega satellite constellations~\cite{9275613}. Next generation satellite communication services are expected to provide seamless connectivity for rural and remote regions with integrated connectivity even in dense urban environments~\cite{1522108}.

The utilization of LEO mega satellite constellations for global communications signifies a paradigm shift from  complete reliance on conventional terrestrial coverage. The trend is to create a hybrid system that combines next generation terrestrial networks and NTN services into a seamless continuum~\cite{9502642}. These new opportunities come with immanent engineering and numerous technical challenges. Some of these challenges are related to network access standardization, inter-satellite connectivity, handover, terrestrial integration, and co-channel interference among many others, all of which impact the network performance. 

The performance of a satellite network is typically evaluated by specialized software platforms that provide tailored simulations. Although simulators provide reasonably reliable results by accounting for delicate details such as antenna gain patterns, Earth terrain, and exact satellite orbits, they do not produce tractable analytical insights. Such insights are critical for capturing the impact of different parameters on the network's performance, making rapid network optimization attainable. Moreover, the complexity of the simulations is aggravated for mega constellations that consist of large numbers of interconnected satellites, increasing the complexity of system optimization. Therefore, there is a need for analytical models that provide trade-offs between real deployments and sophisticated simulation platforms that are characteristically intractable. Analytical modeling that rely on tools from stochastic geometry are extensively deployed in terrestrial networks to capture network performance, and have recently been shown to be viable for mega satellite constellations~\cite{9313025,9177073}.

\begin{figure*}
	\centering
	\includegraphics[width=0.65\linewidth]{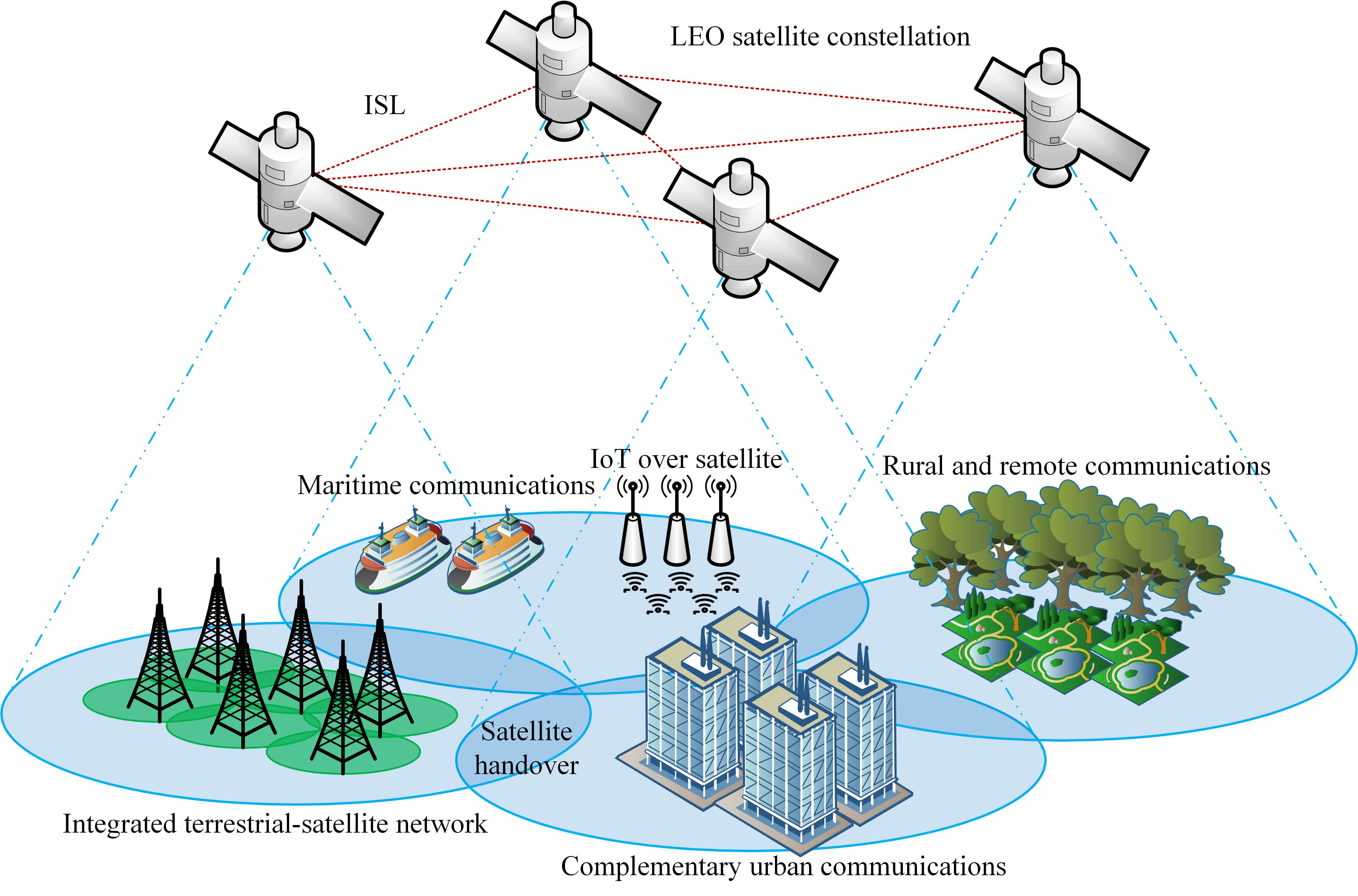}
	\caption{Overview of LEO satellite constellation network showcasing its coverage and prospect opportunities.} 
	\label{Fig_Overview}
\end{figure*}

The four focal contributions of this article are highlighted as follows. The article (i) outlines the benefits of LEO mega satellite constellations and provides insights into the main opportunities, (ii) highlights the challenges that stem from their deployment in next generation networks, (iii) details state-of-the-art analytical performance methods for various network architectures, and (iv) investigates emerging simulation approaches suitable for research and development of mega satellite constellations comprising large numbers of satellites. Fig.~\ref{Fig_Overview} highlights the main features of LEO mega constellations.


\section{Opportunities of Mega Constellations}
\subsection{Fair and Global Coverage}
With the development of 5G and beyond-5G communications, ubiquitous connectivity and high data rates are required to support the envisaged use cases. However, terrestrial services do not necessarily ensure consistent and fair user experience because of difficulties associated with remote region implementation. Satellites emerge as a complementary solution to provide broadband communications beyond urban areas~\cite{1522108}. Assuming the satellite utilizes a sufficiently large beam, a LEO satellite can provide equivalent coverage of 3\% and 12\% of Earth's surface for altitudes of 400 and 2000 km, respectively. As such, rural and remote areas could be efficiently covered to reduce the digital divide. Satellite coverage is also suitable for maritime communications and high-speed moving platforms. One example is providing high-speed reliable wireless connectivity to airplanes or high-speed trains via satellite communication links, which can enable in-cabin Internet services. In addition, maritime connectivity improves safety and vessel traffic efficiency. 
\vspace{-3 mm}
\subsection{IoT over Satellite}
Internet-of-Things (IoT) has gained significant momentum over the past decade, serving many applications that rely on massive wireless sensor networks such as smart cities, transport, remote agriculture, and maritime monitoring. Specifically, low power wide area networks have been largely successful in addressing large area IoT connectivity via terrestrial networks and suitable for battery operated devices. However, many other use-cases require connectivity in rural and remote regions which are difficult to achieve using conventional terrestrial links. Satellite constellations have recently been proposed to support IoT networks~\cite{de2015satellite}, however, many challenges emerge; (i) popular IoT protocols mostly rely on random access techniques resulting in higher co-channel interference generated by the large number of devices competing for spectrum access in the uplink (which is exacerbated by the large coverage area of the satellite), and (ii) IoT devices are typically power limited making the uplink power budget design extremely challenging, even for low orbit satellites. As a result, IoT over satellite protocols, need to implement mechanisms compatible to mitigate co-channel interference with minimal energy budgets.
\vspace{-3 mm}
\subsection{Reduced Latency}
Global coverage via satellites can be realized using geostationary Earth orbit (GEO), medium Earth orbit (MEO) and LEO satellites. Whilst GEO satellite covers a much larger region of the Earth's surface, the typical transmission round trip time (RTT) is over 250 - 350 ms, making them impractical for delay intolerant applications. Similarly, while the altitude of MEO satellites are slightly lower, their RTT is still relatively large. Compared with GEO and MEO based satellite networks, LEO constellations have much lower altitude (typically 400-2000 km). This relative proximity significantly enhances the link budget and reduces the communication latency, e.g., less than 30~ms RTT. However, this proximity also leads to a reduced coverage footprint per satellite, thus stems the need to compensate by deploying larger numbers of satellites per constellation. Low altitudes also leads to reduced launching costs especially with the emergence of small to medium size launchers. Therefore, LEO mega satellite networks have recently attracted intensive interests, such as the networks currently being deployed by OneWeb and Starlink.
\vspace{-2 mm}
\subsection{Multi-user MIMO}
It has been shown that multiple-input multiple-output (MIMO) wireless systems can significantly increase the spectral efficiency by exploiting spatial multiplexing gain. In order to achieve a high spatial multiplexing gain, a low spatial correlation between antennas is required, which can usually be achieved when antennas are sufficiently spaced. For multi-user MIMO, various configurations can be considered to form an antenna array at satellites~\cite{Schwarz19}:
\begin{itemize}
	\item \textbf{Single satellite:} a satellite with multiple antennas.
	\item \textbf{Collocated satellites:} multiple satellites (each equipped with a single antenna) in the same orbital plane.
	\item \textbf{Multiple satellites:} each equipped with single a antenna in different orbital plane.
\end{itemize}
When multiple satellites form an antenna array, synchronization through inter-satellite links becomes critical as antennas (or satellites) are distributed. On the other hand, the number of possible antennas is limited when a single satellite is used in multi-user MIMO scheme. Nevertheless, multi-user MIMO is expected to be a key enabler in making satellite communications competitive in terms of increasing spectral efficiency and supporting high-throughput applications.
\vspace{-2 mm}
\subsection{Recent Development Examples}
\subsubsection{OneWeb}
OneWeb is a global communications network, currently in the process of deploying a constellation of LEO satellites. The initial deployment of 650 satellites will provide connectivity to fixed, maritime and aviation user terminals for governments, businesses, and communities who are beyond the reach of fiber networks. As well as the satellites themselves, the system includes a global terrestrial network of interconnected ground stations, points-of-presence, and operations centers. This provides affordable, high throughput, and low-latency communication services which will enable IoT systems of the future, and a pathway to 5G for everyone, everywhere.
\subsubsection{SpaceX}
SpaceX's Starlink is another global communications network that is providing Internet services to a large portion of Earth's surface. Starlink's constellation is developing to consist of over 4,000 satellites, at altitudes between 540 and 570 km, divided into five different shells coupled with inter satellite links (ISL) a feature enabling satellites to communication between one another.
\subsubsection{Amazon}
Amazon announced the launch of their constellation, dubbed as Keuper, to provide Internet service to Earth except its two poles. The highly anticipated satellite constellation is expected to hold a total of 3,236 satellites at an altitude of 630 km. 

\section{Challenges of Mega Constellations}

\subsection{Radio Channel for LEO Orbits}
LEO satellite links provide enhanced link power budgets relative to higher altitude satellites, enhancing their appeal in next generation NTNs. However, there are several challenges emerging in the satellite-to-ground radio channel related to the fast relative movement of LEO satellites. Unlike conventional geostationary links, LEO satellites have lower orbital periods. For instance, a LEO satellite located at an altitude of 400 km takes approximately 1.5 hours to complete its orbit around Earth, whereas satellites located at 2000 km take up to 2.1 hours. This means that a satellite dwells from only 6.25 to 12.5 minutes within the ground user's field-of-view~\cite{9422812}.

Since LEO satellites move at a remarkably fast pace within the field-of-view window, large variations in the ground user elevation angle occur, ensuing large fluctuations in the power link budget. The free-space path-loss, affected by the distance between satellite and ground user, ranges from its minimum when the satellite is located at the user's zenith to a maximum value at the local horizon. Moreover, the probability of achieving line-of-sight communications for LEO links is heavily dependent on the elevation angle~\cite{9257490}. Another implication to the velocities at which LEO satellites move, are the resulting high Doppler frequency shifts that occur regardless of the ground terminal velocities. In addition, higher spectral regions are currently being explored such as the Q/V-band and above for their potential usage in mega LEO constellations. These bands offer immense spectral vacancy, however, at the cost of increased adverse channel conditions including increased atmospheric absorption and susceptibility to extreme fluctuations. As such, robust channel prediction will provide satellite systems with the capability in adapting to large variations in the radio channel. 
\vspace{-2 mm}
\subsection{Co-channel Interference}
The radio spectrum is a finite resource shared by a myriad of wireless transmitters. Due to the growing numbers of wireless technologies constantly competing over the scarce spectrum, the reuse of spectral resources has become more vital to achieve better spectral efficiency by enhance multiple access technologies. Consequently, with large numbers of transmissions trying to jointly access the frequency bands, higher levels of co-channel interference are generated. Thus, next generation satellite access systems need to utilize intelligent distributed multi-access mechanisms/scheduling to enable the coexistence of transmissions. Moreover, co-channel interference caused by higher altitude satellites such as GEO need to be addressed. 

\begin{figure*}
	\captionsetup[subfigure]{labelformat=empty}
	\begin{subfigure}{0.33\textwidth}
		\centering
		\caption{Random Constellation}
		\includegraphics[width=\linewidth]{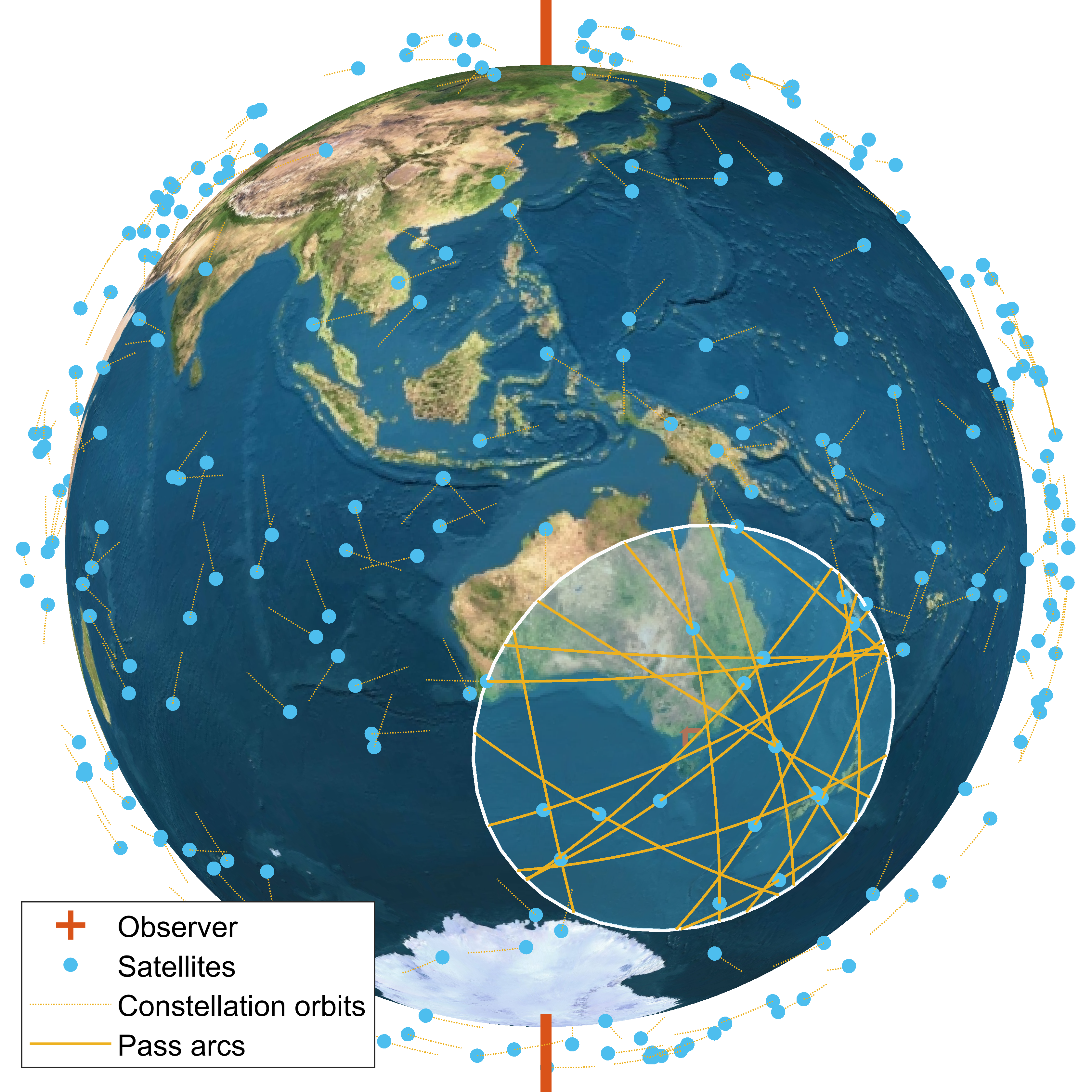}
	\end{subfigure}
	\begin{subfigure}{0.33\textwidth}
		\centering
		\caption{Walker Delta Constellation}
		\includegraphics[width=\linewidth]{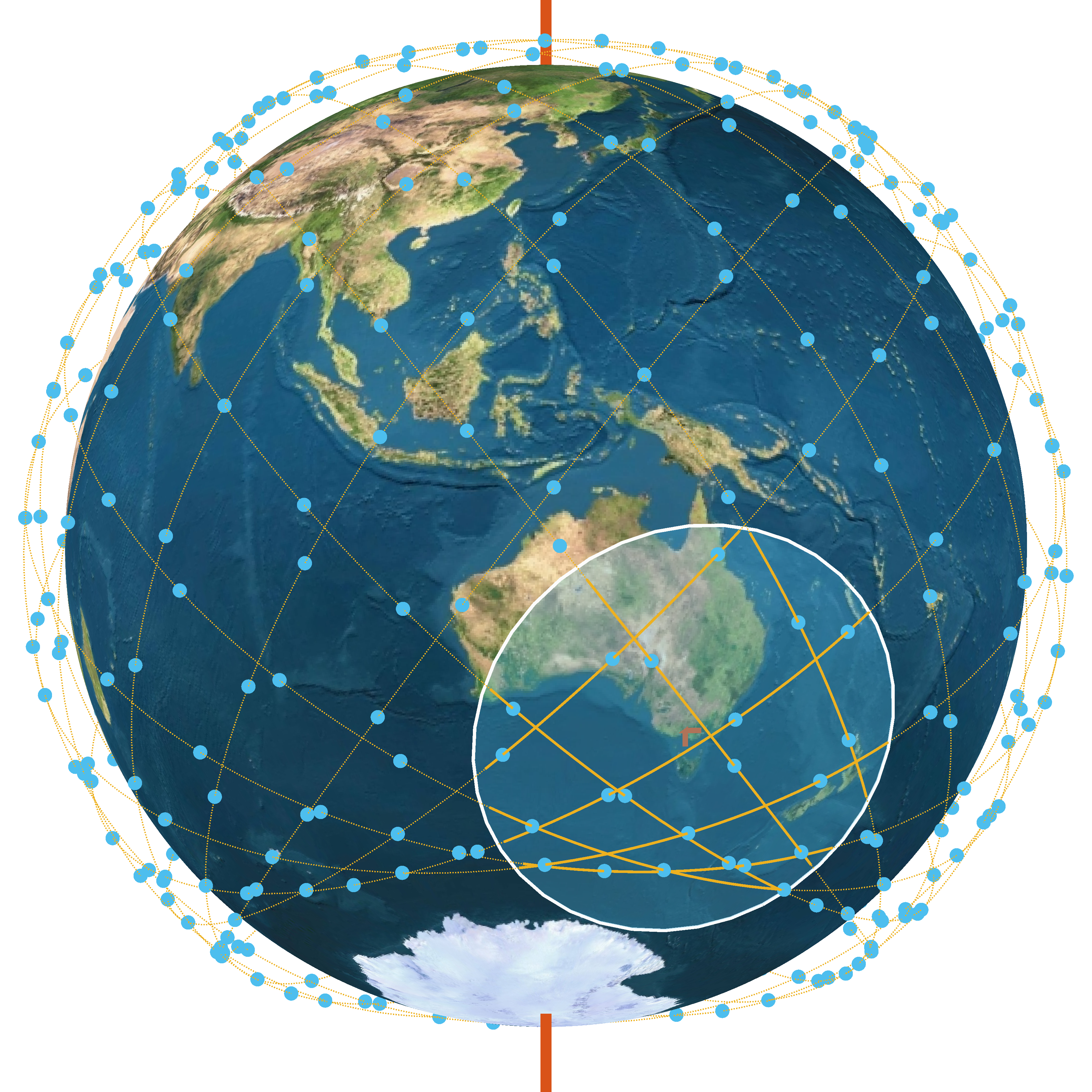}
	\end{subfigure}
	\begin{subfigure}{.33\textwidth}
		\centering
		\caption{Walker Star Constellation}
		\includegraphics[width=\linewidth]{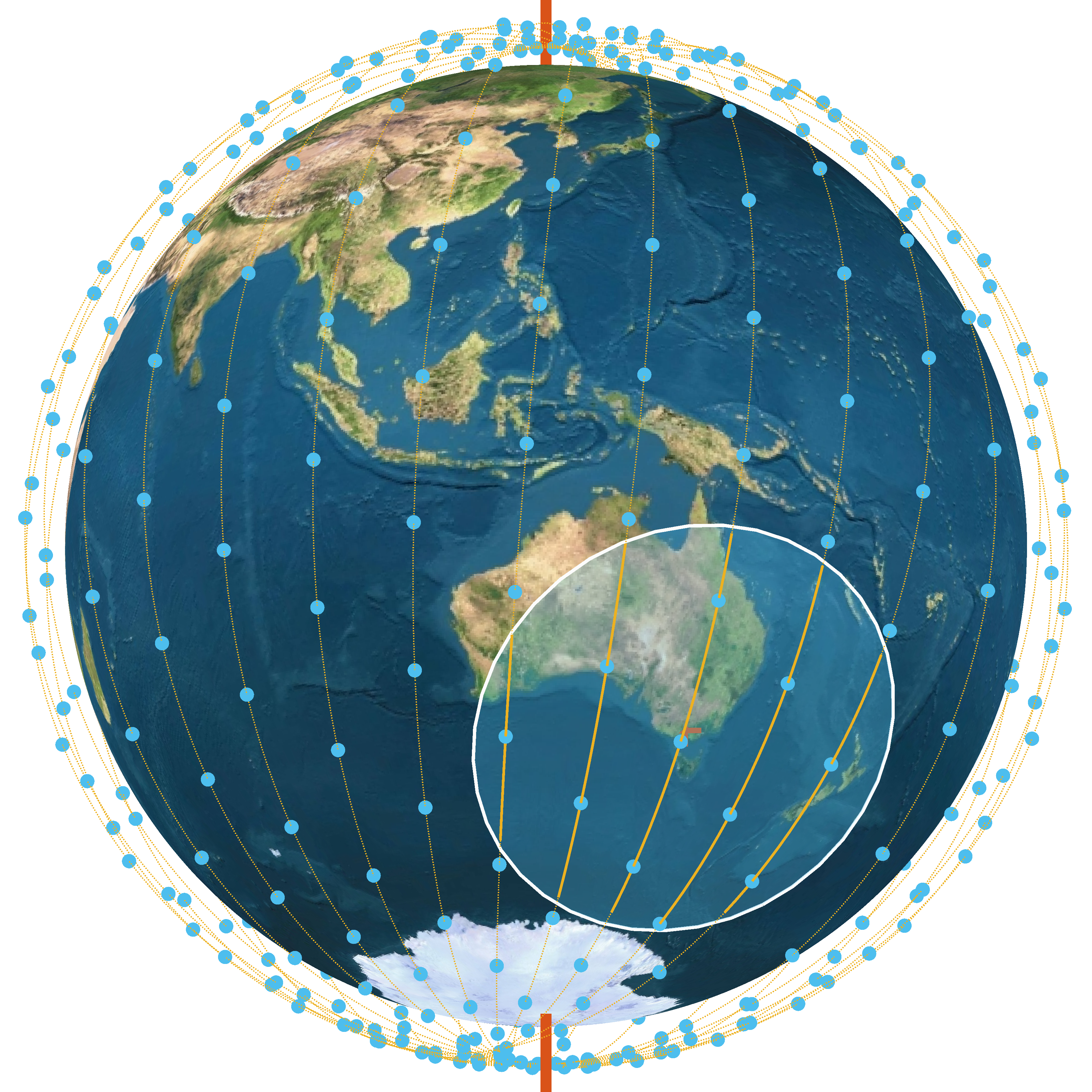}
	\end{subfigure}
	\caption{Time snap showcasing the satellite geometric distribution for random, Walker-delta, and Walker-star constellations.} 
	\label{Fig_LEOConstellation}
\end{figure*}

For the downlink, the large numbers of satellites deployed in a LEO constellation (above 1,000 satellites per constellation), are expected to share the same spectral band. Therefore, multiple satellites within the ground user's local dome\footnote{The local dome is the section of the constellation sphere within which satellites are able to connect to the ground user, see Fig.~\ref{Fig_LEOConstellation}.} of a ground user can potentially interfere with each other. The co-channel interference (within a given constellation) can be significantly reduced by employing intelligent beamforming, fractional frequency reuse, and inter-satellite interference mitigation protocols. However, spectral sharing between different constellations or between satellite and terrestrial services, is a much harder problem due to the inherent lack of coordination. Therefore, careful spectrum management is required in order to regulate the parameters of network access technologies.

Similarly, for the uplink, the interference generated by the sheer number of ground users can cause significant performance degradation in case the spectral bands are shared with terrestrial services. This issue is exacerbated because of the reduced path-loss towards LEO satellites. Moreover, with a large coverage footprint of the satellites (compared with its terrestrial counterpart), a larger uplink coordination region is required. Similar to downlink methods, several interference mitigation schemes can be employed. An additional barrier is the global nature of satellite services which require the harmonization of multiple standards and different regulation bodies around the world.
\vspace{-4 mm}
\subsection{Integration with Terrestrial Networks}
Beyond 5G communications, integrated terrestrial-satellite networks (ITSN) are expected to cooperatively include both terrestrial and satellite networks to combine their benefits. In such hybrid scheme, high data rates by terrestrial networks and large coverage footprint of satellites are combined to produce a seamless continuum~\cite{9502642}. Major challenges in such an integrated scheme are three folds; (i) cross satellite-terrestrial radio resource coordination, (ii) cross platform handover, and (iii) performance modeling, validation, and optimization. 3GPP is actively looking into the integration of NTNs under future 5G (and beyond) releases, where several technical reports are addressing different aspects of such integration from the physical layer to architecture integration options~\cite{5GNTN}.

\vspace{-3 mm}
\subsection{Network Access Standardization}
Historically, there are limited examples of network access standardization for satellite services, with geostationary satellite commonly operating as a bent-pipe architecture with dedicated modems being paired between user terminals and gateways. While components are interchangeable or generic (parabolic antennas for example), interoperability between modem vendors is not typical. An example of standardization that was attempted in mobile satellite services for GEO-Mobile Radio Interface 1 and 2 to substantially reuse terrestrial mobile telephony standards. Although both have been standardized by the European Telecommunications Standards Institute (ETSI), they are not supported by many systems to enable economies of scale. With the development of these standards stalled, standardization of satellite services are currently being pioneered by 3GPP~\cite{5GNTN}. In specific, standardization of network access for LEO mega constellations has the additional challenges compared relative to GEO as follows; (i) orbital parameters include altitude, inclination, number of orbital planes and satellites, (ii) physical technical limitations including satellite size weight and power, (iii) different business requirements for different operators, and (iv) regulatory restrictions including frequency band and market access.

These differences impact the technology choices and by extension the ability to standardize the network access in a number of ways. For instance, unlike GEO satellites, LEO constellations can operate on a variety of altitudes, orbital planes, and inclinations, which in turn produces significant variations between systems in terms of delay and path-loss. Optimizing a standard for one system architecture may inhibit or reduce the performance of others. Some operators’ business requirements may be tailored towards commercial and government services requiring controllable Quality-of-Service while others may be tailored to consumer best-effort services. System architectures are currently somewhat fluid and under constant revision. For example, SpaceX’s Starlink system parameters have changed multiple times prior to and during its implementation with the most recent FCC submission proposing two different future architectures based on potential availability of launch vehicles. The competitive nature of early system deployments and the continual updating of system architectures are significant impediments to satellite network access standardization.

\begin{figure*}
	\captionsetup[subfigure]{labelformat=empty}
	\begin{subfigure}{0.5\linewidth}
		\centering
		\includegraphics[width=0.9\linewidth]{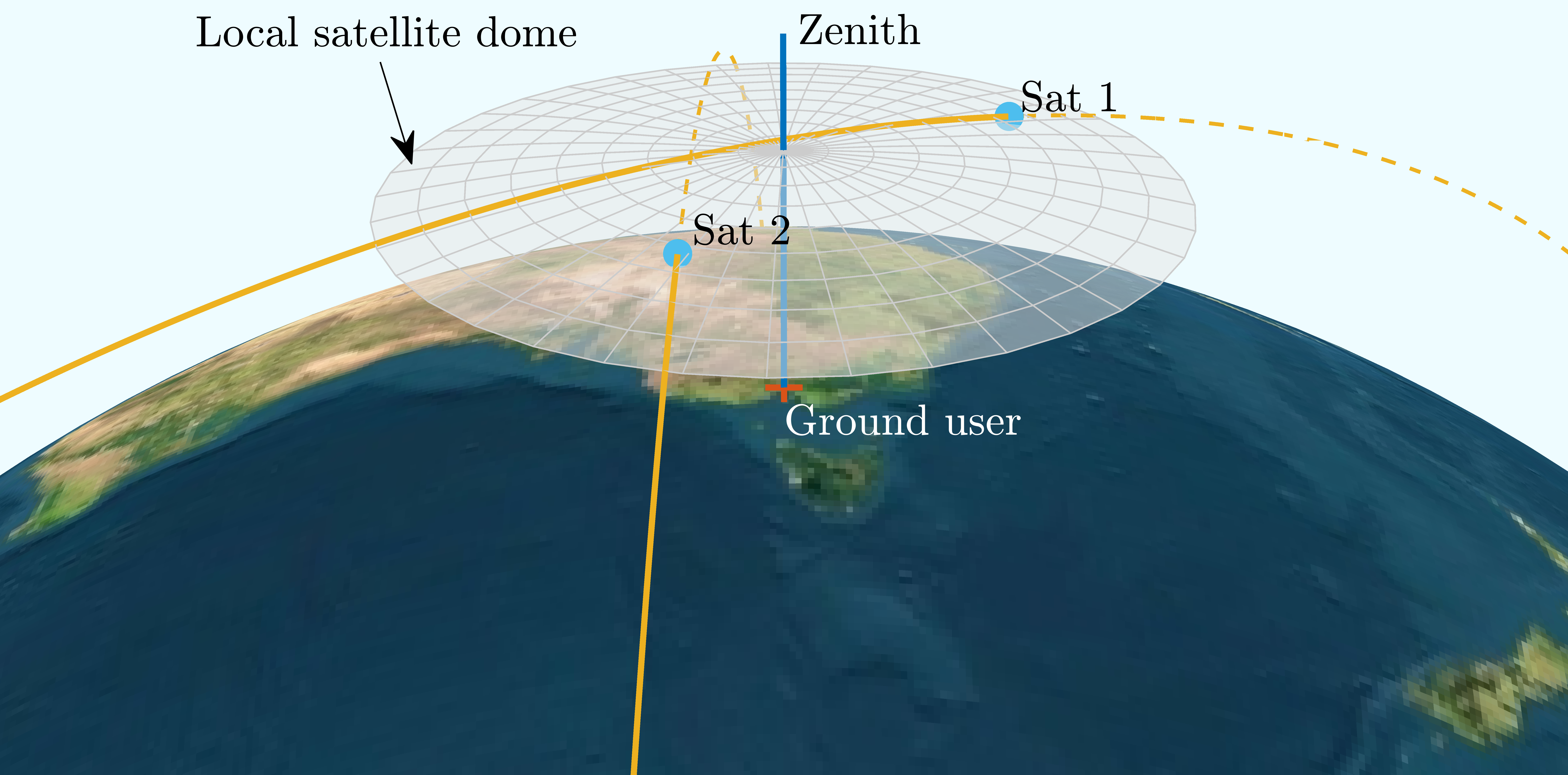}
		\caption{User-centric model}
	\end{subfigure}
	\begin{subfigure}{0.5\linewidth}
		\centering
		\includegraphics[width=0.9\linewidth]{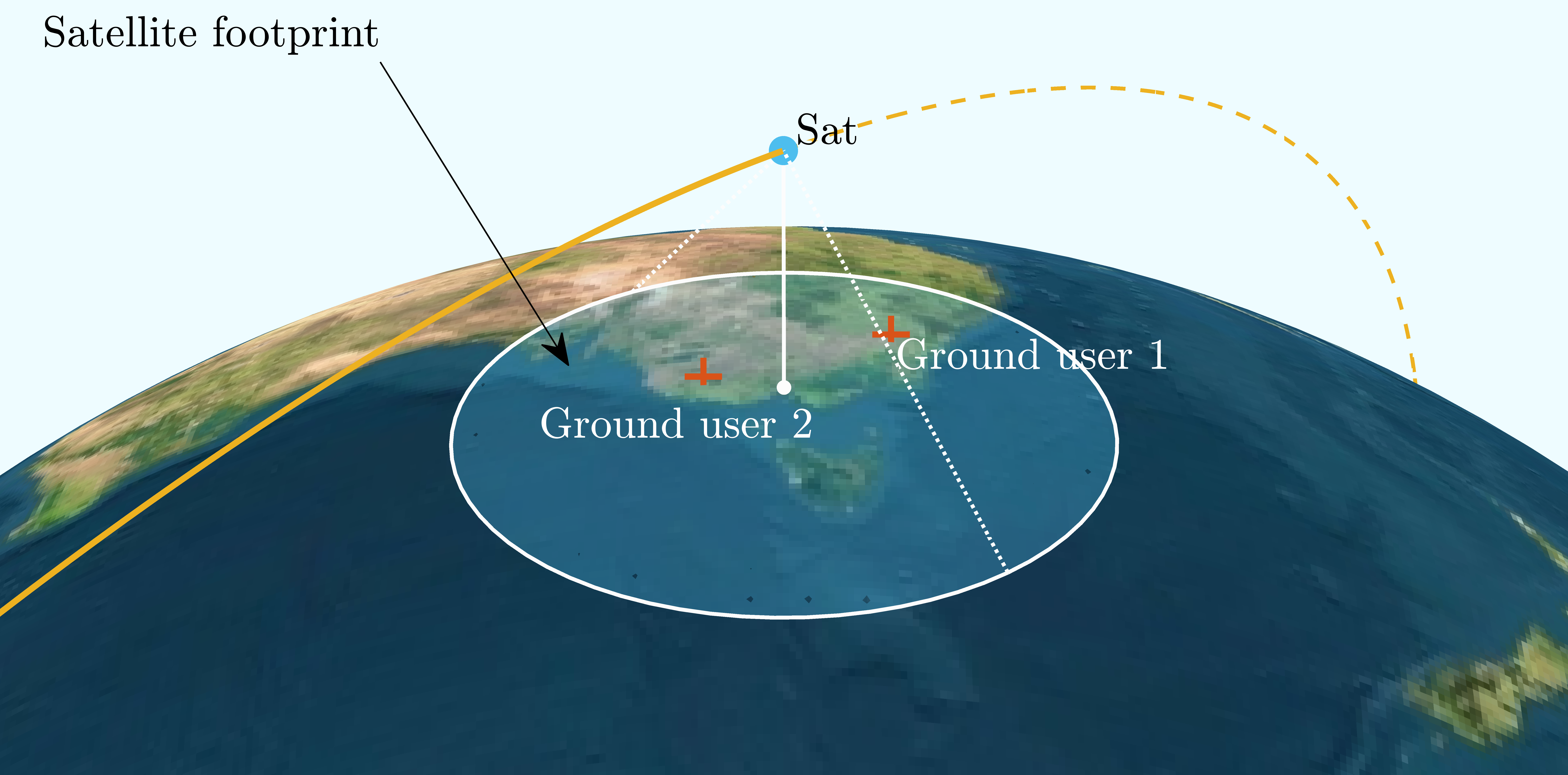}
		\caption{Satellite-centric model}
	\end{subfigure}
	\caption{Difference between user and satellite-centric analysis in downlink and uplink analysis, respectively. Interference is generated by satellites within the local dome in downlink, whereas by ground users within the footprint in uplink analysis.} 
	\label{Fig_DomeVsFootrpint}
\end{figure*}
\vspace{-2 mm}
\subsection{Inter-Satellite Connectivity}
ISLs are desirable to realize satellite networks, particularly in LEO constellations with large numbers of satellites. ISLs enable data communication, interconnection and information relay between satellites. Without ISLs, mega satellite networks will require massive numbers of gateways to provide real-time global coverage. Moreover, ISLs are highly desirable for long-distance low-latency communications, where the use of LEO satellite network with ISLs can reduce the latency compared with optical fiber-based terrestrial networks. Hence, ISLs can enable latency-sensitive services, such as financial applications (e.g. low latency connections between financial centers). Moreover, the trend of onboard processing and increased intelligence in satellites also requires ISLs to realize automated network management. Both the radio frequency and the optical bands can be explored for ISLs~\cite{sodnik2010optical}. Compared with RF-based solutions, whilst the use of optical spectrum for ISLs has the requirement of precise beam pointing, it has key advantages of broader bandwidth, lower power consumption and lower interference. Therefore, the use of optical ISLs can achieve higher data rate and improved signal-to-noise ratio (SNR). Whilst ISLs have been extensively studied and developed, there are a number of critical challenges to be tackled. One key challenge is the handover between different satellites, particularly when there are large numbers of satellites in different orbital altitudes. This issue becomes even worse with optical ISLs, due to the narrow beam size that requires precise acquisition, tracking and pointing (ATP) and frequent handovers. In addition, the dynamic property of LEO-based satellite networks also leads to network topology changes, and hence an increased signalling overhead.

\vspace{-3 mm}
\subsection{Handover}
LEO satellites orbit at very high speed, thus dwelling only for few minutes in the effective antenna range of the user. This is a quite short period, where within it, the network is required to execute a complex handover process in order to maintain uninterrupted user's session. Handover procedures, are quite mature in terrestrial networks with continuing improvement road-map of 5G towards more-reliable and lower-latency handovers, in such networks the handover procedure is typically triggered due to the user mobility. To the contrary, satellite handovers occur due to the fast relative movement of the satellites coupled with a continuously changing network topology. Several challenges are linked to frequent handovers, e.g., (i) increased signaling overhead, (ii) more frequent re-routing of traffic (and thus increased latency), and (iii) more mechanical wear in mechanically steerable antennas. As such, it is very important to characterize the handover rate, example of such analysis is provided in~\cite{Handover}, the findings in this paper give a rapid understanding the handover timing without resorting to resource-intensive simulations.

\section{Performance of Mega Constellations}
\subsection{Constellation Geometric Modeling}


Two popular constellation patterns are currently being used by recent deployments, and these patters are (i) Walker-delta, adopted by Starlink, and (ii) Walker-star, adopted by OneWeb. Such highly regular patterns cannot be analytically investigated and rely on exhaustive simulation approaches which are not tractable. On the other hand, stochastic geometry approaches have recently been extended to LEO mega constellation networks. Unlike terrestrial networks, LEO satellite constellations are in a continuous orbital motion. In order to attain analytic tractability, random satellite constellation model is proposed to leverage the inherent features of Poisson point process (PPP). The satellites in the random model follow random circular orbits with a uniformly random starting point. These starting points are randomly and uniformly distributed on a spherical surface according to a Binomial point process (BPP)~\cite{9177073}. However, from the local observer point of view, the distribution of the satellite points in the local satellite dome can be approximated to a PPP because it is a much smaller subset of the constellation sphere~\cite{9313025}. As such, the PPP model enables tractable statistical properties of the contact angle which facilitates analytic interference and coverage analysis. Indeed, this random model does not replicate well-designed practical dense satellite constellations, however it is proven that random constellations can provide close insights and approximations to practical deployments~\cite{9177073,9313025}, or can rapidly capture the bounds of the performance behavior. A visual comparison between the three different constellations is provided in Fig.~\ref{Fig_LEOConstellation}.

\vspace{-3 mm}
\subsection{Satellite Network Performance}
The analytical performance of a network is usually gauged by the average success probability of the radio link (or the radio packet). The success of a satellite communication link is predicated on the radio channel condition and the utilized media access control (MAC). Furthermore, the performance analysis of the network differs for downlink and uplink communication systems. Fig.~\ref{Fig_DomeVsFootrpint} illustrates the difference between adopting user-centric and satellite-centric models for the downlink and uplink analysis, respectively.

\subsubsection{Downlink}
The performance of downlink communications is obtained by adopting a user-centric model, where a typical user is assumed to be located randomly on Earth's surface. Any active satellite that is within the user's local satellite dome is capable of establishing a communication link with the respective user. The area of the local satellite dome monotonically increases with the maximum zenith angle. The convention is to allocate the nearest satellite as the serving satellite. Consequently, all the other active satellites within the dome are considered to be potential interferers~\cite{9313025}. 

\subsubsection{Uplink}
The uplink performance, on the other hand, is evaluated using a satellite-centric model, where a typical ground user is randomly located in its footprint. As such, ground users that are located within the same footprint are considered as potential interferers. The interference level will significantly depend on the multiple access scheme and ground user density. However, due to the expected large numbers of ground users, analytic approaches leverage the low variations in the instantaneous interference power, and hence the interference can be adequately represented with its average~\cite{9509510}.

\vspace{-3 mm}
\subsection{Hybrid ITSN models}
Analytical modeling for ITSN provides designers with rough estimates towards the required network infrastructure both the dimensioning of the terrestrial and satellite components. Moreover, an inherent benefit of modeling the performance of satellite constellations using stochastic geometry is the presence of well-established terrestrial network models, whereby the performance for ITSN can be consequently obtained by combining the two models. Two prominant ITSN configurations can be considered: 
\subsubsection{Relay ITSN}
utilizes terrestrial base stations to relay transmissions between satellites and ground users, to mitigate the signal degradation which occurs due to shadowing. By combining a BPP model to represent the satellites in the constellation with an independent PPP for the terrestrial base stations, the enhanced coverage in rural environments can be analytically obtained~\cite{9218989}.

\subsubsection{Heterogeneous ITSN}
enables satellites and terrestrial base stations to jointly access the same spectral resources based on coordinated radio technology, whether centrally coordinated or distributed. Since ground users may be served by both networks simultaneously, the performance is dependent on both channels. A system that deploys two independent PPPs for both the satellite constellation and terrestrial base stations can capture the impact of the constellation size and base station density on the link performance~\cite{9509510}.

\vspace{-3 mm}
\subsection{Pass-duration}
Pass-duration is the interval where the satellite is \emph{available} for establishing a reliable communication link with a given ground user. Predicting satellite passes and tracking non-geosynchronous orbits is an important functionality for almost all satellite-based communications, sensing, and positioning systems. The prediction algorithms in these systems provide an accurate estimation of the moment at which the satellite enters the observer's field-of-view, including the coordinates vector along the pass-duration. The simplified general perturbations (SGP) models are commonly applied for tracking near-Earth satellites based on a set of parameters describing the orbit at a given point in time. Thus, clearly with increasing computational power, it is not a difficult task to obtain the statistical distribution of the pass-duration for a certain satellite constellation. However, such calculations are not tractable and could not be heuristically inverted, i.e., they do not provide the ability to obtain the orbital parameters based on desired pass-duration statistics. Estimating the pass duration of a constellation allows better understanding of multiple practical aspects including, for example, (i) the required signaling overhead for handovers between the satellites, (ii) the maximum un-interrupted communication session than can be established between a satellite and a user, and (iii) the required wake up duration in energy-constrained IoT devices, so it will upload the needed sensor information and receive downlink commands if needed. The analytic study in~\cite{9422812} provides a tractable approach for obtaining the mean and the distribution of satellite pass-duration that is solely dependent on the constellation altitude and the antenna beamwidth. 
\begin{figure}
	\centering
	\includegraphics[width=\linewidth]{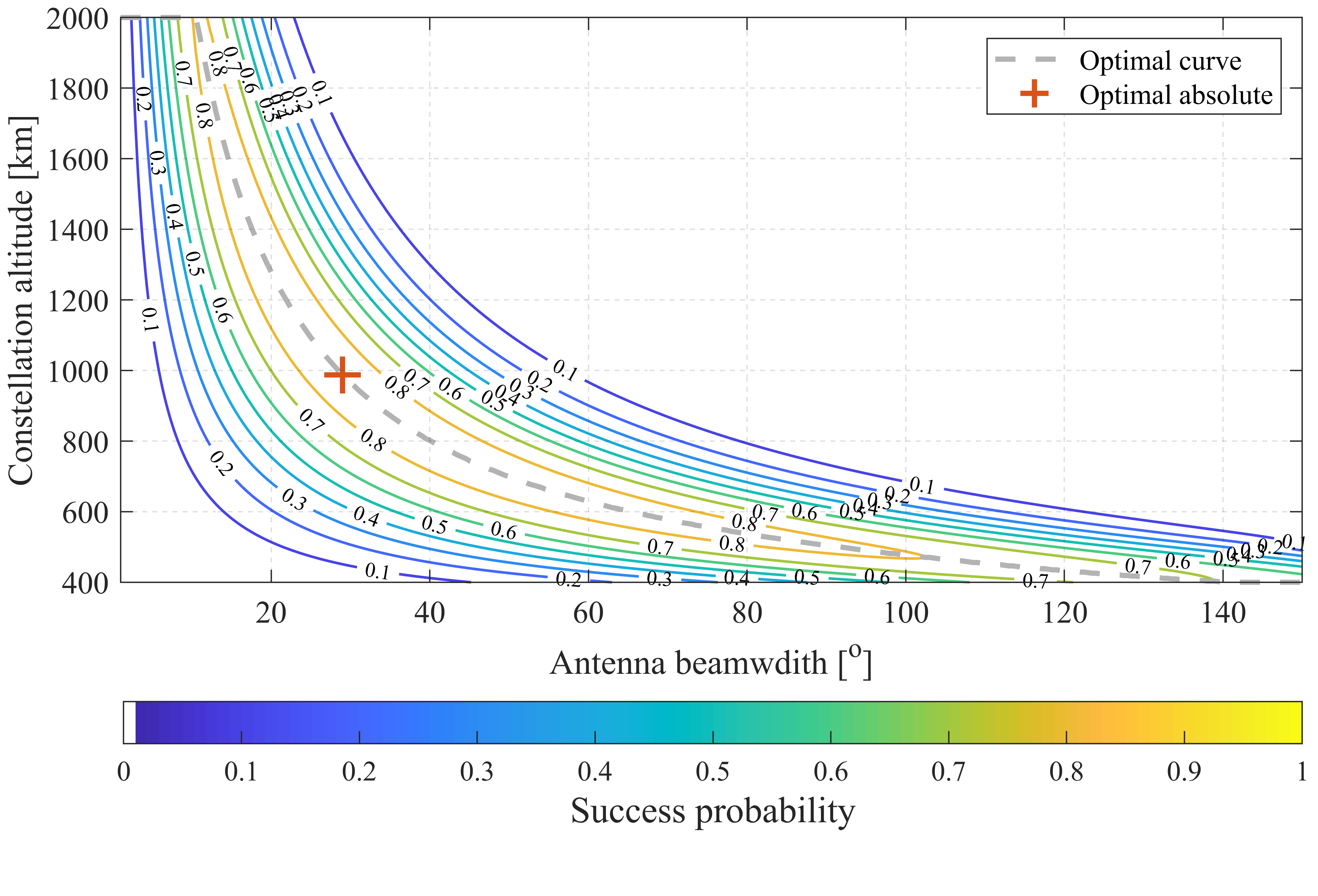}
	\caption{Analytical beamwidth and constellation altitude joint optimization for satellite uplink communication.}
	\label{Fig_Optimization}
\end{figure}
\vspace{-4 mm}
\subsection{Optimization}
The primary benefit of employing analytic modeling based on stochastic geometry is the ability to conduct rapid performance optimization. This optimization can either be achieved analytically (using differentiation) or using well-established numeric algorithms. For example, the optimization for mega LEO constellations can be performed based on maximizing the success probability by tuning key network parameters. Among the many important parameters that impact the performance is the number of satellites in a constellation, their altitude above the mean sea level~\cite{9390220}, and the antenna beamwidth that the satellite or ground users utilize in the network. Moreover, optimization approaches can be extended to hybrid ITSN for jointly tuning terrestrial and satellite network parameters~\cite{9509510}. An example of joint optimization of satellite beamwidth and constellation altitude is illustrated in Fig.~\ref{Fig_Optimization}.
\vspace{-2mm}
\section{Simulation Approaches}
\begin{figure}
	\centering
	\includegraphics[width=\linewidth]{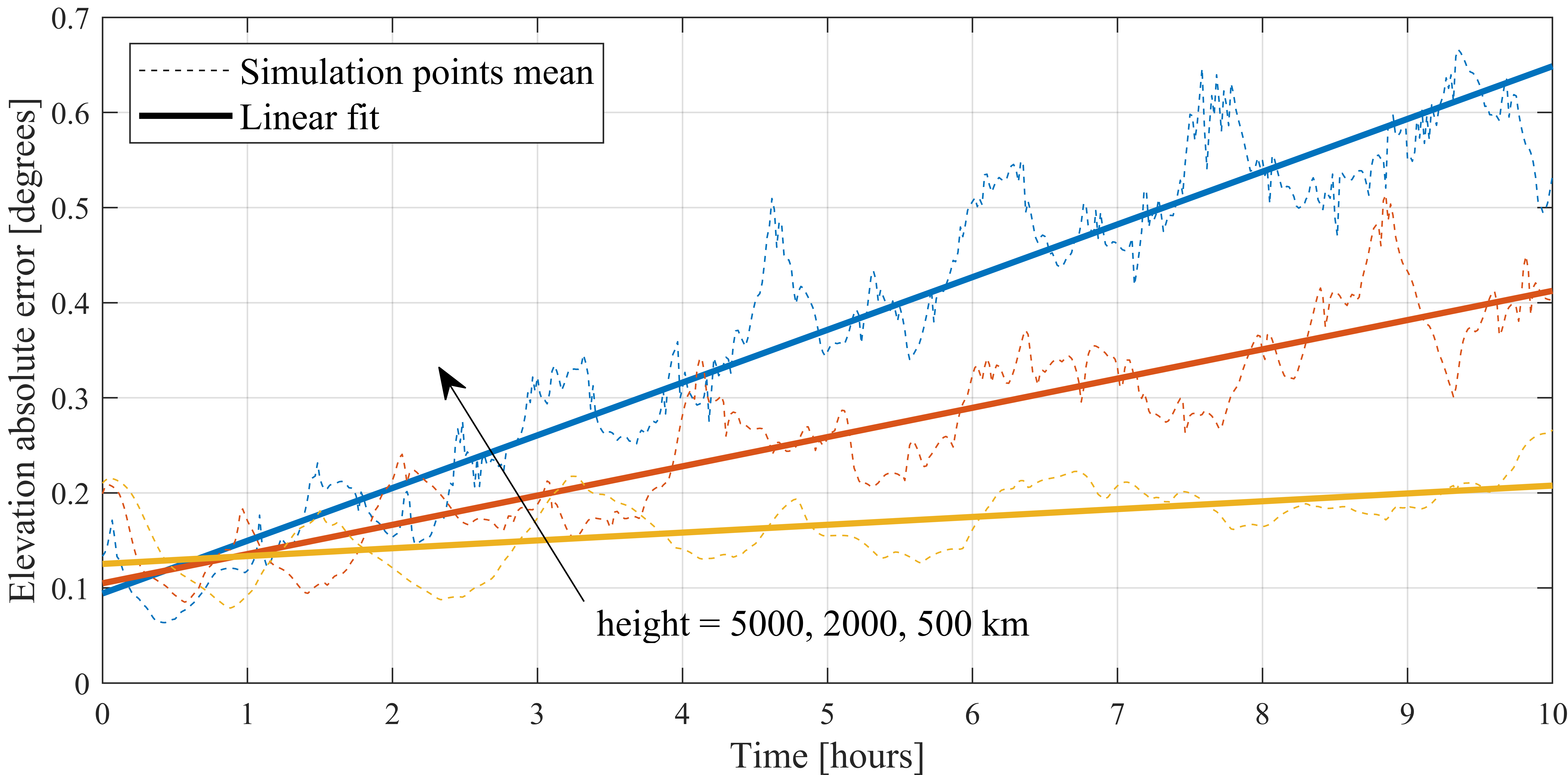}
	\caption{Elevation absolute error comparison between two-body propagator and SGP4 for different constellation heights.}
	\label{Fig_Simulation}
\end{figure}

Orbital propagators allow the prediction of satellites' positions at certain time instance given a known position and velocity at another time instance. There are many orbital propagation algorithms with various levels of accuracy and applications, however we can generally classify orbital propagator into two (i) analytic and (ii) numeric propagators. Numeric propagator utilizes numeric integration to predict the satellite orbit and provides high-fidelity outcome, hence this is more suitable for satellite operation and control. Analytic propagators use simplified closed-form approximated solutions to provide rapid orbit estimation at the cost of lower fidelity, hence this is more suitable for rapid wireless parameter optimization and to perform multiple what-if scenarios. Analytic propagators are also suitable for operation and control. Under the analytic propagator group, we mention three important methods that are also supported by common scripting tools:
\begin{itemize}
	\item \textbf{Two-body propagator} is the simplest propagator based on the Keplerian orbital model. It is quick to generate orbits using this model and can provide rapid results, making it especially useful for optimizing LEO massive satellite constellations.
	\item \textbf{SGP4} or the Simplified General Perturbations propagator suitable for operation and control and does not necessarily add enhanced accuracy in the design phase.
	\item \textbf{SDP4} or the Simplified Simplified Deep Space Perturbations propagator, similar to SGP4, but also takes into account lunar and solar gravity perturbations, giving higher fidelity for orbits above LEO.
\end{itemize}
The absolute difference between the simplest two-body propagator and SGP4 for three different constellation altitudes is illustrated in Fig.~\ref{Fig_Simulation}. Note that within the first few hours, the difference is too small to affect the radio conditions.
\vspace{-2mm}
\section{Conclusion}
In the face of a globally accruing digital divide, there is a need to modify the existing network infrastructure that enables reimagining key network components that makeup conventional networks and reassessing networks' performance metrics. This article provides an overview of the emerging LEO mega satellite constellations for next generation wireless communication networks. It outlines some of the opportunities that position LEO constellations as a promising solution to attaining global coverage. Moreover, it highlight the challenges that emerge in the hybridization of networks and the technical aspects involved in their deployment. Additionally, a comprehensive analysis of LEO mega satellite constellations that address the analytical and simulation-based performance modeling is presented. 


\ifCLASSOPTIONcaptionsoff
\newpage
\fi
\bibliographystyle{IEEEtran}
\bibliography{Satellite_Magazine.bib}

\vspace{-13 mm}
\begin{IEEEbiographynophoto}{Bassel Al Homssi}
	is a Research Fellow at both RMIT University and Deakin University, Melbourne, Australia. Dr Al Homssi is currently working on multiple research projects related to satellite networks and MIMO communications for next generation wireless systems.
\end{IEEEbiographynophoto}
\vspace{-13 mm}
\begin{IEEEbiographynophoto}{Akram Al-Hourani} is a Senior Lecturer and Telecommunication Program Manager at the School of Engineering, RMIT University. He has extensive industry/government engagement as a chief investigator in multiple research projects related to Satellite Communications, The Internet-of-Things (IoT), and Smart Cities.
\end{IEEEbiographynophoto}
\vspace{-13 mm}
\begin{IEEEbiographynophoto}{Ke Wang} is a Senior Lecturer and Australian Research Council (ARC) DECRA Fellow at RMIT University. He is also the Program Manager of Telecommunications Engineering. His research interests mainly include optical and wireless communications and convergence, machine learning in telecommunications, and integrated opto-electrronic devices and circuits. 
\end{IEEEbiographynophoto}
\vspace{-13 mm}
\begin{IEEEbiographynophoto}{Phillip Conder}
	is a Senior Research Fellow at RMIT University. His experience includes research on Spectrum Management, terrestrial and satellite wireless communications in Government, Defence and Startup environments.
\end{IEEEbiographynophoto}
\vspace{-13 mm}
\begin{IEEEbiographynophoto}{Sithamparanathan Kandeepan}
	is a Professor in Telecommunications Engineering, RMIT University. His research interests are in wireless and satellite communications, he currently leads the satellite communications program for the Space Industry Hub, works with Smart Satellite Cooperative Research Centre and leads the Wireless Innovation Lab (WiLAB).
\end{IEEEbiographynophoto}
\vspace{-13 mm}
\begin{IEEEbiographynophoto}{Jinho Choi}
	is a Professor at Deakin University, Melbourne, Australia. His research interests include statistical
	signal processing and wireless communications. He authored two books, Adaptive and Iterative Signal Processing in Communications and Optimal Combining and Detection, published by Cambridge University Press, and currently serves as an Editor for IEEE Transactions on Communications.
\end{IEEEbiographynophoto}
\vspace{-13 mm}
\begin{IEEEbiographynophoto}{Ben Allen}
	is Director of System Modelling at OneWeb. He has been researching radio communications for several decades, spanning both academia and industry. He holds a PhD in communications engineering from the University of Bristol and recently completed a Royal Society Industry Fellowship in collaboration with the University of Oxford.  in Wireless Communications. He holds a Royal Society Industry Fellowship. 
\end{IEEEbiographynophoto}
\vspace{-13 mm}
\begin{IEEEbiographynophoto}{Ben Moores}
	is Satellite Communications Systems Engineer working in the Advanced Architecture Group at OneWeb. He previously focused on digital satellite payloads, working on both the operator and provider sides. He holds a MEng from Cambridge.
\end{IEEEbiographynophoto}




\end{document}